\documentclass[aps,showpacs,showkeys,superscriptaddress]{revtex4}

\usepackage{graphicx}
\usepackage{makeidx}
\usepackage[]{hyperref}
\usepackage{amsmath}

\usepackage{amssymb,epsfig,graphics}
\usepackage{amscd}
\usepackage{verbatim}

\usepackage{amsmath}
\usepackage{amsthm}
\usepackage{amsfonts}

\begin{document}
\title{Spin--orbit interaction in the graphitic nanocone}

\author{R. Pincak}\email{pincak@saske.sk}
\affiliation{Institute of Experimental Physics, Slovak Academy of Sciences,
Watsonova 47,043 53 Kosice, Slovak Republic}
\affiliation{Bogoliubov Laboratory of Theoretical Physics, Joint
Institute for Nuclear Research, 141980 Dubna, Moscow region, Russia}

\author{J. Smotlacha}\email{smota@centrum.cz}
\affiliation{Bogoliubov Laboratory of Theoretical Physics, Joint
Institute for Nuclear Research, 141980 Dubna, Moscow region, Russia}
\affiliation{Faculty of Nuclear Sciences and Physical Engineering, Czech Technical University, Brehova 7, 110 00 Prague,
Czech Republic}

\author{M.Pudlak}\email{pudlak@saske.sk}
\affiliation{Institute of Experimental Physics, Slovak Academy of
Sciences, Watsonova 47,043 53 Kosice, Slovak Republic}

\date{\today}

\begin{abstract}
The Hamiltonian for nanocones with curvature--induced spin--orbit
coupling have been derived. The effect of curvature--induced
spin--orbit coupling on the electronic properties of graphitic
nanocones is considered. Energy spectra for different numbers of the
pentagonal defects in the tip of the nanocones are calculated. It was
shown that the spin--orbit interaction considerably affects the local
density of states of the graphitic nanocone. This influence depends on the
number of defects present at the tip of the nanocone. This property could be applied in atomic force microscopy for the construction of the probing tip.
\end{abstract}



\maketitle

\section{Introduction}\

The spin--orbit interaction in graphene is supposed to be weak, due
to the low atomic number of carbon. spin--orbit coupling (SOC) in
graphene has an intrinsic part, completely determined from the
symmetry properties of the honeycomb lattice. The strength of this
intrinsic spin--orbit coupling is rather small, due to weakness of
the intra--atomic spin--orbit coupling of carbon. Because of the symmetry
of the honeycomb lattice this intrinsic spin--orbit coupling vanishes for
hopping between neighboring atoms \cite{kane}. To get the
contribution from this kind of spin--orbit coupling we have to go to the
next order in the hopping. In this paper we work with
the tight--binding approximation where we take into account only the
hopping between nearest neighboring atoms.

In a curved graphene sheet where the symmetry of honeycomb lattice
is broken there is a possibility of curvature--induced spin--orbit
coupling. A consistent approach to introduce this kind of SOC has
been developed by Ando \cite{Ando}. The experimental evidence for
this kind of spin--orbit coupling was reported by Kuemmeth et al. \cite{Kuemmeth}. It was demonstrated that in clean nanotubes the
spin and orbital motion of electrons are coupled. In this work the authors measured the values of spin--orbit coupling in carbon nanotubes at various values of the magnetic field strength. It was revealed that the symmetry in electron--hole
spectrum is broken. This can be caused by spin--orbit coupling.

In \cite{Fang} the influence of SOC on the Kondo effect in
carbon nanotube quantum dots was investigated by Fang et al. The results indicate
that the spin--orbit coupling significantly changes the low--energy
Kondo physics in carbon nanotube quantum dots. Recently, Steele et al.
\cite{Steele} have reported the large spin--orbit coupling in carbon
nanotubes. It turns out that the spin--orbit coupling could be
significantly enlarged by the nonzero curvature of the nanoparticle
surface \cite{Brataas,Jeong,Valle}. Energy spectra and transport
properties of armchair nanotubes with curvature--induced spin--orbit
interactions were investigated by Pichugin et al \cite{PPN}. It was
reported that due to spin--orbit coupling an armchair nanotube can
serve in some energy range as an spin filter. To understand clearly
quantum phenomena in carbon nanoparticles the spin--orbit coupling
has to be included to describe their electronic properties. The
spin--orbit coupling could also be important in nanocones due to their curved surface.

In this paper, we derive the effective Hamiltonian for the
graphitic nanocone with spin--orbit coupling induced by curvature.
The structure of the paper is as follows. In Sec. II, we introduce
an explicit formula for the eigenspectrum of the Hamiltonian with
full curvature--induced spin--orbit coupling in a carbon nanocone. The
solution is derived in the Appendices. In Sec. III, we present numerical
results for all of our calculations. The main conclusions are
summarized in Sec. IV.

\section{Dirac equation for curvature--induced spin--orbit coupling}\

\begin{figure}[htbp]
\includegraphics[width=45mm]{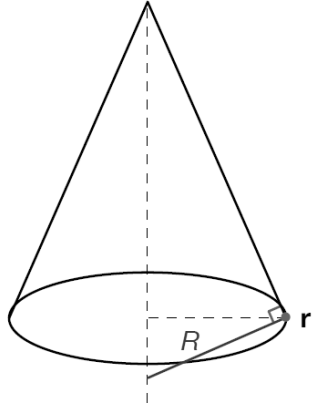}
\caption{The notation of the distances in the nanocone.}\label{konus}
\end{figure}

The Hamiltonian has been derived following the method of Ando
\cite{Ando}. Adapting the derivation of Hamiltonian in \cite{sitenko}, we introduce the curvature--induced spin--orbit
interaction on the nanocone. We start with the Hamiltonian for the
nanoconical surface without the spin--orbit coupling and
pseudopotential \cite{sitenko,Pincak}. Due to the rotational
symmetry of the nanocone, we choose the radial and angular
coordinates $r, \varphi$. Here, we will often use the coordinate
$R$. It is the distance between the point $\bold{r}$ on the surface
and the intersection of the conical axis with the line perpendicular
to surface at point $\bold{r}$ (Fig. \ref{konus}). It satisfies
$R=\frac{(1-\eta)r}{\sqrt{\eta(2-\eta)}}$, where $\eta=N_d/6$ is the
Frank index depending on $N_d$, the number of defects in the
nanoconical tip.

The Hamiltonian has the form
\begin{equation}\hat{H}=\left(\begin{array}{cc}H_1 & 0\\0 & H_{-1}\end{array}\right)\hspace{1cm}
H_s={\rm i}\hbar v\left\{\tau^y\partial_r-\tau^xr^{-1}\left[(1-\eta)^{-1}\left(s\partial_{\varphi}-\frac{3}{2}{\rm i}\eta\right)-
\frac{1}{2}\tau^z\right]\right\},\end{equation}
where $\tau^x, \tau^y, \tau^z$ are the Pauli matrices, $s=\pm 1$ denotes the value of the $K$ spin.

Now, we will supply the terms corresponding to the spin--orbit interaction. It means that we perform the substitutions
\begin{equation}\partial_r\,\rightarrow\,\partial_r-\frac{\delta\gamma'}{4\gamma}b^{\varphi}_{\varphi}\sigma_x(\vec{r}),
\hspace{1cm}{\rm i}\partial_{\varphi}\,\rightarrow\,{\rm
i}\partial_{\varphi}+s(1-\eta)A_y\sigma_y.\end{equation} Here,
$\sigma^x, \sigma^y, \sigma^z$ are the Pauli matrices corresponding
to spin of electrons and we use the linear combination
$\sigma_{x}(\vec{r})=\sigma^{x}\cos\varphi-\sigma^{z}\sin\varphi$.
The curvature of the surface is included in the curvature tensor
$b^{\varphi}_{\varphi}=\frac{1}{R}$ and in the parameter
$A_y=s\frac{2\delta p}{(1-\eta)}\sqrt{\eta(2-\eta)}$, which depends
on the Frank index. The other parameters are described in
\cite{Ando} and they have the following meaning: the hopping
integrals $\gamma =-\sqrt{3}V_{pp}^{\pi}a/2$,
$\gamma^{'}=\sqrt{3}(V_{pp}^{\sigma}-V_{pp}^{\pi})a/2$, where $a$ is
the length of the primitive translation vector ($a=\sqrt{3}d\simeq
2.46 {\AA}$, $d$ is the distance between atoms in the unit
cell); $V_{pp}^{\sigma}$ and $V_{pp}^{\pi}$ are the transfer
integrals for the $\sigma$ and $\pi$ orbitals, respectively, in flat 2D graphite. Next, $p=1-3\gamma^{'}/8\gamma$, $\delta =
\frac{1}{3}\frac{\Delta}{\epsilon_{\pi\sigma}}$, where
\begin{equation}
\Delta = i\frac{3\hbar}{4m^{2}c^{2}}\langle x|\frac{\partial
V}{\partial x} p_{y}-\frac{\partial V}{\partial y} p_{x}|y\rangle
\end{equation}
($V$ is the atomic potential) and
$\epsilon_{\pi\sigma}=\epsilon_{2p}^{\pi}-\epsilon_{2p}^\sigma $.
The energy $\epsilon_{2p}^{\sigma}$ is the energy of
$\sigma$--orbitals that are localized between carbon atoms. The
energy $\epsilon_{2p}^{\pi}$ is the energy of $\pi$--orbitals that
are directed perpendicular to the nanotubular surface. The following
values of the parameters are chosen: $\delta$ is of the order
between $10^{-3}$ and $10^{-2}$, $\frac{\gamma'}{\gamma}\sim
\frac{8}{3}$, $p\sim 0.1$, $V_{pp}^{\pi}\approx -3$ eV and
$V_{pp}^{\sigma}\approx 5 $ eV, $p\approx 0.1$ and $\delta \approx
0.01$\cite{Ando}. Now, we will do the transformation
\begin{equation}\hat{H'}=e^{{\rm i}\frac{\sigma_y}{2}\varphi}\hat{H}e^{-{\rm i}\frac{\sigma_y}{2}\varphi}\end{equation}
With the aid of this transformation we describe the motion of an
electron in the local coordinate frame which moves with the electron on
the nanocone surface. Now we get
\begin{equation}\hat{H'}=\hbar v\left(\begin{array}{cc}0 & \partial_r-{\rm i}s\frac{1}{r(1-\eta)}\partial_{\varphi}-\frac{1}{2r}
-{\rm i}\frac{1}{r}\xi_x\sigma_x-\frac{3}{2}\frac{\eta}{(1-\eta)r}-\frac{\xi_y\sigma_y}{r}\\
-\partial_r+{\rm i}s\frac{1}{r(1-\eta)}\partial_{\varphi}-\frac{1}{2r}
+{\rm i}\frac{1}{r}\xi_x\sigma_x-\frac{3}{2}\frac{\eta}{(1-\eta)r}-\frac{\xi_y\sigma_y}{r}& 0\end{array}\right),\end{equation}
where the parameters $\xi_x, \xi_y$ describe the strength of the spin--orbit interaction:
\begin{equation}\xi_x=\frac{\delta\gamma'\sqrt{\eta(2-\eta)}}{4(1-\eta)\gamma},\hspace{1cm}                                \xi_y=A_{y}+\frac{1}{2(1-\eta)}.\end{equation}

Now, we are solving the equation
\begin{equation}\label{DirEq}\hat{H'}\psi(r,\varphi)=E\psi(r,\varphi),\end{equation}
where, considering the rotational symmetry of the solution, we do the factorization
\begin{equation}\psi(r,\varphi)=e^{{\rm i}j\varphi}\left(\begin{array}{c}f_{j\uparrow}(r)\\
f_{j\downarrow}(r)\\g_{j\uparrow}(r)\\g_{j\downarrow}(r)\end{array}\right).\end{equation}
Then, after the substitution of this expression into (\ref{DirEq}) and performing the differentiations with respect to $\varphi$, we have
\begin{equation}\label{syst}\left(\begin{array}{cccc}0 & 0 & \partial_r+\frac{F}{r} & -\frac{\rm i}{r}C\\0 & 0 & -\frac{\rm i}{r}D &
\partial_r+\frac{F}{r}\\ -\partial_r+\frac{F-1}{r} & \frac{\rm i}{r}D & 0 & 0\\
\frac{\rm i}{r}C & -\partial_r+\frac{F-1}{r} & 0 & 0\end{array}\right)\left(\begin{array}{c}f_{j\uparrow}(r)\\
f_{j\downarrow}(r)\\g_{j\uparrow}(r)\\g_{j\downarrow}(r)\end{array}\right)=E\left(\begin{array}{c}f_{j\uparrow}(r)\\
f_{j\downarrow}(r)\\g_{j\uparrow}(r)\\g_{j\downarrow}(r)\end{array}\right),\end{equation}
where
\begin{equation}F=\frac{sj}{1-\eta}-\frac{3}{2}\frac{\eta}{1-\eta}+\frac{1}{2},\hspace{1cm}C=\xi_x-\xi_y,
\hspace{1cm}D=\xi_x+\xi_y.\end{equation}
The parameter $s=\pm 1$. The solution of (\ref{syst}) is given in the Appendix.

\section{Electronic properties of the conical nanostructure}\

\begin{figure}[htbp]
\includegraphics[width=180mm]{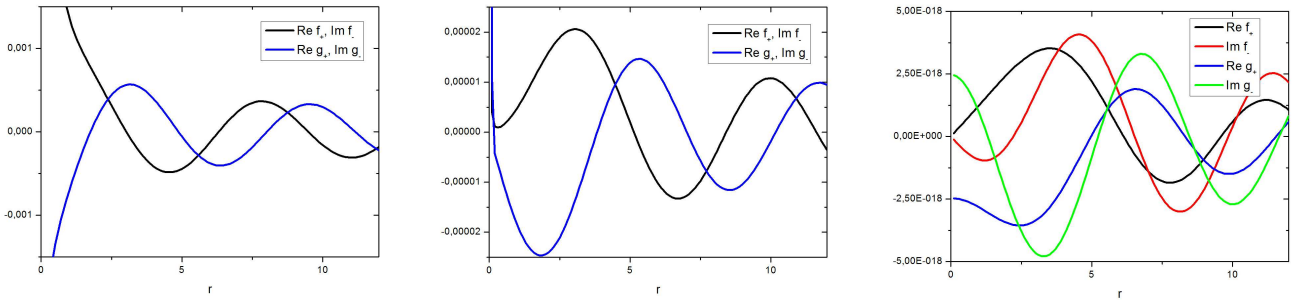}
\caption{Solution of the system (\ref{syst}) for the case of
$N_{d}=1$ (left), $N_{d}=2$ (middle), $N_{d}=3$ (right) and
$E=1$.}\label{ReIm}
\end{figure}

For $\beta=1$ and $E=1$, we see the solution of (\ref{syst}) in Fig.
\ref{ReIm}. The signs $\uparrow, \downarrow$ in the indices are
replaced here by the signs $+,-$, respectively. The graphs for
Re$\,f_{\downarrow}$, Re$\,g_{\downarrow}$, Im$\,f_{\uparrow}$ and
Im$\,g_{\uparrow}$ are missing -- they provide the zero solution. We
can see that for the cases of 1 and 2 defects, the modules of
$f_{\downarrow}$, $f_{\uparrow}$ and $g_{\downarrow}$,
$g_{\uparrow}$, respectively, coincide -- the existing effect of the
spin--orbit interaction is still not strong enough to split the
appropriate components of the wave--function.

This solution has a form similar to Bessel functions of the first
or the second kind ($J_j$ or $Y_j$). This correspondence is derived
in Appendix B for the case of $N_{d}=1$ defect and $j=2$. For other
values of the energies or the parameter $j$, this occurrence will be
similar.

For the normalized case, given energy and value of $j$, the local density of states ($LDoS$) is defined as
\begin{equation}LDoS(E,r)=|f_{j,E\uparrow}(r)|^2+|f_{j,E\downarrow}(r)|^2+|g_{j,E\uparrow}(r)|^2+|g_{j,E\downarrow}(r)|^2.
\end{equation}

\begin{figure}[htbp]
\includegraphics[width=160mm]{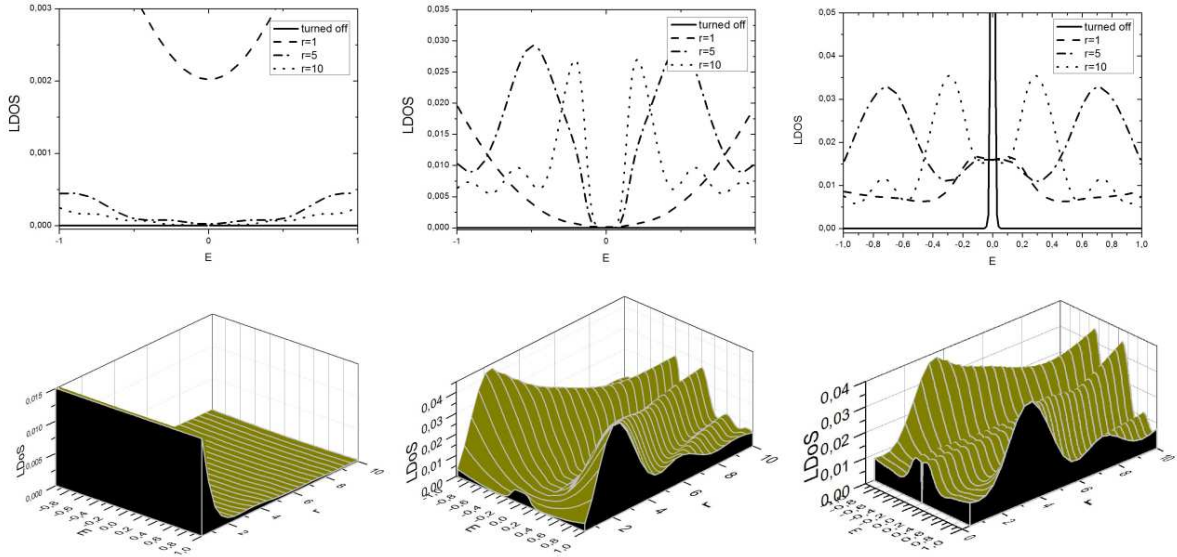}
\caption{2D and 3D (bottom) graphs of the local density of states
with and without (turned off) spin--orbital interaction for different
distances $r$ from the tip. $N_{d}=1$ (left), $N_{d}=2$ (middle) and
$N_{d}=3$ (right).}\label{LDoS_d}
\end{figure}

The numerical results are depicted in Fig. \ref{LDoS_d}. The
solution without the spin--orbit interaction is the turned off case. For this case, the constants $C$ and $D$ would be equal to zero and the system (\ref{syst}) would be a 4--dimensional analogy of the 2--dimensional case without the spin--orbit interaction (described by Pincak et al. in \cite{Pincak}) and with the exclusion of the effect of the pseudopotential. This
evidence can also be derived from the above plots in Fig.
\ref{ReIm}, where the modules of the first and second or the third
and fourth component, respectively, are not split unless the number of the
defects in the conical tip exceeds 2.

\section{Conclusion}\

An effective mass Hamiltonian was derived for a carbon nanocone in the presence of curvature--induced spin--orbit interaction. Within our
approach, we solved analytically the eigenvalue problem for the
effective mass Hamiltonian for electrons on the curved surface with the
spin--orbit interaction. In particular, we obtained explicit
expressions for a low--energy spectrum and eigenstates of carbon
nanocones. The $LDoS$ in the graphitic nanocone near the tip in the
case of spin--orbit interactions were computed numerically. These
findings have been used to analyze electronic properties of carbon
nanocones with curvature--induced SOC at different limits. We see that the spin–orbit interaction considerably affects the local density of states of the graphic nanocone; the higher the number of defects, the bigger the effect. One of the reason is that the more defects
present near the tip, the bigger the curvature of the nanocone in the
vicinity of the tip, and so also the bigger the effect of the imposed
spin--orbit interaction. The localization of the electrons as shown
in Fig. \ref{LDoS_d} makes the graphitic nanocone a possible
candidate for the construction of a new type of scanning probe in atomic force microscopy \cite{chenchen, carbon,nanomgn1,
nanomgn2}.\\

ACKNOWLEDGEMENTS --- The work was supported by the Science and Technology Assistance Agency under Contract No. APVV-0171-10, VEGA Grant No. 2/0037/13 and Ministry
of Education Agency for Structural Funds of EU in frame of project
 26220120021, 26220120033 and 26110230061. R. Pincak would like to thank the
 TH division in CERN for hospitality.

\appendix

\section{Solution of the Dirac equation}\

We want to find the solution of (\ref{syst}) in the form
\begin{equation}\label{solF}f_{j\uparrow}(r)=e^{\frac{\alpha}{r}+\beta r}\sum\limits_{k=0}^{\infty}a_kr^{\xi+k},\hspace{1cm}
f_{j\downarrow}(r)=e^{\frac{\alpha}{r}+\beta r}\sum\limits_{k=0}^{\infty}b_kr^{\xi+k},\end{equation}
\begin{equation}\label{solG}g_{j\uparrow}(r)=e^{\frac{\alpha}{r}+\beta r}\sum\limits_{k=0}^{\infty}c_kr^{\xi_1+k},\hspace{1cm}
g_{j\downarrow}(r)=e^{\frac{\alpha}{r}+\beta r}\sum\limits_{k=0}^{\infty}d_kr^{\xi_1+k}.\end{equation}
After the substitution, we get $\xi=\xi_1-2$ and
\begin{equation}\label{eqA3}-\alpha c_0=Ea_0,\hspace{1cm}-\alpha c_1+\xi_1c_0+Fc_0-{\rm i}Cd_0=Ea_1,\hspace{1cm}-\alpha d_0=Eb_0,\hspace{1cm}-\alpha d_1+\xi_1d_0+Fd_0-{\rm i}Dc_0=Eb_1,\end{equation}
\begin{equation}\alpha a_0=0,\hspace{1cm}\alpha a_1+\xi a_0+(F-1)a_0+{\rm i}Db_0=0,\hspace{1cm}\alpha b_0=0,\hspace{1cm}\alpha b_1-\xi b_0+(F-1)b_0+{\rm i}Ca_0=0,\end{equation}
\begin{equation}\alpha a_2-\beta a_0+(F-\xi-2)a_1+{\rm i}Db_1=0,\hspace{1cm}\alpha b_2-\beta b_0+(F-\xi-2)b_1+{\rm i}Ca_1=0,\end{equation}
\begin{equation}\alpha a_3-\beta a_1+(F-\xi-3)a_2+{\rm i}Db_2=0,\hspace{1cm}\alpha b_3-\beta b_1+(F-\xi-3)b_2+{\rm i}Ca_2=0.\end{equation}
For the other indices, we get the system of recurrence equations
\begin{equation}-\alpha c_k+\beta c_{k-2}+(F+\xi_1+k-1)c_{k-1}-{\rm i}Cd_{k-1}=Ea_k,\hspace{1cm}-\alpha d_k+\beta d_{k-2}+(F+\xi_1+k-1)d_{k-1}-{\rm i}Dc_{k-1}=Eb_k,\end{equation}
\begin{equation}-\alpha a_k+\beta a_{k-2}-(F-\xi_1+2-k)a_{k-1}-{\rm i}Db_{k-1}=-Ec_{k-4},\hspace{1cm}-\alpha b_k+\beta b_{k-2}-(F-\xi_1+2-k)b_{k-1}-{\rm i}Ca_{k-1}=-Ed_{k-4}.\end{equation}
If we suppose that $\alpha\neq0$, we get the zero solution. So for the nontrivial solution $\alpha=0$ and as follows from the first and the third equation in (\ref{eqA3}), in this case the coefficients $a_0$ and $b_0$ must be also zero. Then, from the system
\begin{equation}(F-\xi_1)a_1+{\rm i}Db_1=0,\hspace{1cm}(F-\xi_1)b_1+{\rm i}Ca_1=0\end{equation}
follows:
\begin{equation}\xi_1=F\pm{\rm i}\sqrt{CD},\hspace{1cm}b_1=\pm\sqrt{\frac{C}{D}}a_1.\end{equation}
From the system
\begin{equation}(F+\xi_1)c_0-{\rm i}Cd_0=Ea_1,\hspace{1cm}(F+\xi_1)d_0-{\rm i}Dc_0=Eb_1,\end{equation}
we get
\begin{equation}c_0=\frac{(F+\xi_1)a_1+{\rm i}Cb_1}{(F+\xi_1)^2+CD}E,\hspace{1cm}d_0=\frac{Eb_1+{\rm i}Dc_0}{F+\xi_1}.\end{equation}
And from the system
\begin{equation}(F-\xi_1-1)a_2+{\rm i}Db_2=\beta a_1,\hspace{1cm}(F-\xi_1-1)b_2+{\rm i}Ca_2=\beta b_1,\end{equation}
follows
\begin{equation}b_2=\beta\frac{(F-\xi_1-1)b_1-{\rm i}Ca_1}{(F-\xi_1-1)^2+CD},\hspace{1cm}
a_2=\frac{\beta a_1-{\rm i}Db_2}{F-\xi_1-1}.\end{equation}
Here, $\beta$ is a free parameter. The following coefficients we get from the recurrence equations.\\

Now excluding the functions $f_{j\uparrow}, f_{j\downarrow}$ with
the help of the first two equations in (\ref{syst}), we get a
simplified system
\begin{equation}\label{simplsyst}-r^2g_{j\uparrow}''(r)+(2F-1)rg_{j\uparrow}'(r)+
\left[F(F-2)+D^2\right]g_{j\uparrow}(r)+{\rm i}r(C+D)g_{j\downarrow}'(r)-{\rm i}(C-D)Fg_{j\downarrow}(r)=E^2r^2g_{j\uparrow}(r),\end{equation}
\begin{equation}{\rm i}r(C+D)g_{j\uparrow}'(r)+{\rm i}(CF-D)g_{j\uparrow}(r)-{\rm i}r(F-1)D g_{j\uparrow}(r)
-r^2g_{j\downarrow}''(r)-rg_{j\downarrow}'(r)+(C^2+F^2)g_{j\downarrow}(r)=E^2r^2g_{j\downarrow}(r).\end{equation}
To solve the problem an iteration method is used. For this purpose,
we divide the components of the solution into the real and the
imaginary part:
\begin{equation}g_{j\uparrow}={\rm Re}\,g_{j\uparrow}+{\rm i}\,{\rm Im}\,g_{j\uparrow},\hspace{1cm}
g_{j\downarrow}={\rm Re}\,g_{j\downarrow}+{\rm i}\,{\rm
Im}\,g_{j\downarrow}.\end{equation}

Now, for a given $j$, we denote
\begin{equation}{\rm Re}\,g_{j\uparrow}=G_1,\hspace{1cm}{\rm Im}\,g_{j\downarrow}=G_2\end{equation}
and if we suppose that the conditions
\begin{equation}\label{fnull}{\rm Im}\,f_{j\uparrow}={\rm Re}\,f_{j\downarrow}={\rm Im}\,g_{j\uparrow}={\rm Re}\,g_{j\downarrow}=0.\end{equation}
are satisfied for the analytical solution, for the nonzero
components of $g_{j\uparrow}, g_{j\downarrow}$ we have
\begin{equation}-r^2G_1''(r)+(2F-1)rG_1'(r)+\left[F(F-2)+D^2\right]G_1(r)-r(C+D)G_2'(r)+(C-D)F G_2(r)=E^2r^2G_1(r),\end{equation}
\begin{equation}r(C+D)G_1'(r)+(CF-D)G_1(r)-r(F-1)D G_1(r)
-r^2G_2''(r)-rG_2'(r)+(C^2+F^2)G_2(r)=E^2r^2G_2(r).\end{equation}
This system can be rewritten into the form
\begin{equation}\label{inhom1}r^2G_1''(r)-(2F-1)rG_1'(r)+\left[E^2r^2-F(F-2)-D^2\right]G_1(r)=\mathfrak{D}_1(G_2(r)),\end{equation}
\begin{equation}\label{inhom2}r^2G_2''(r)+rG_2'(r)+\left[E^2r^2-C^2-F^2\right]G_2(r)=\mathfrak{D}_2(G_1(r)),\end{equation}
where $\mathfrak{D}_1, \mathfrak{D}_2$ on the right--hand side denote the differential operators. If we exclude them, we get the homogeneous parts of the system. For the second equation it gives the Bessel equation, for the first equation it gives a Bessel--like equation. It is not a problem to find a solution for this homogeneous system. We can try to find a particular solution for the inhomogeneous system in this way: let $G_{1-}^{(0)}, G_{1+}^{(0)}$ and $G_{2-}^{(0)}, G_{2+}^{(0)}$, respectively, denote the solution for the homogeneous system (actually, $G_{2-}^{(0)}, G_{2+}^{(0)}$ are the Bessel functions), then the particular solutions could be searched with the help of the method of variation of the constants, i.e. in the form
\begin{equation}G_1(r)=C_{1-}(r)G_{1-}^{(0)}(r)+C_{1+}(r)G_{1+}^{(0)}(r),\hspace{5mm}
G_2(r)=C_{2-}(r)G_{2-}^{(0)}(r)+C_{2+}(r)G_{2+}^{(0)}(r).\end{equation}
Of course, there is the question of in which form the functions
$G_1(r), G_2(r)$ would be substituted into the arguments of the
operators $\mathfrak{D}_1, \mathfrak{D}_2$. In the first step, we
could make a statement $C_{2-}(r)=C_{2+}(r)=1$ and substitute
$G_2(r)$ into the right--hand side of (\ref{inhom1}). In this way, we
get the form of $C_{1-}(r), C_{1+}(r)$ and we can substitute
$G_1(r)$ into the right--hand side of (\ref{inhom2}) acquiring more
precise values of $C_{2-}(r), C_{2+}(r)$. This procedure can be
repeated unless we achieve the required precision. We make a suggestion that the solution
of (\ref{syst}) has a form of the Bessel--like functions. This estimate will be verified
in the next section by comparison of the coefficients $a_n, b_n, c_n, d_n$ with
the coefficients corresponding to the Taylor series of the Bessel functions.\\

\section{Verification of the similarity between the solution and Bessel functions}

\begin{figure}[htbp]
\includegraphics[width=150mm]{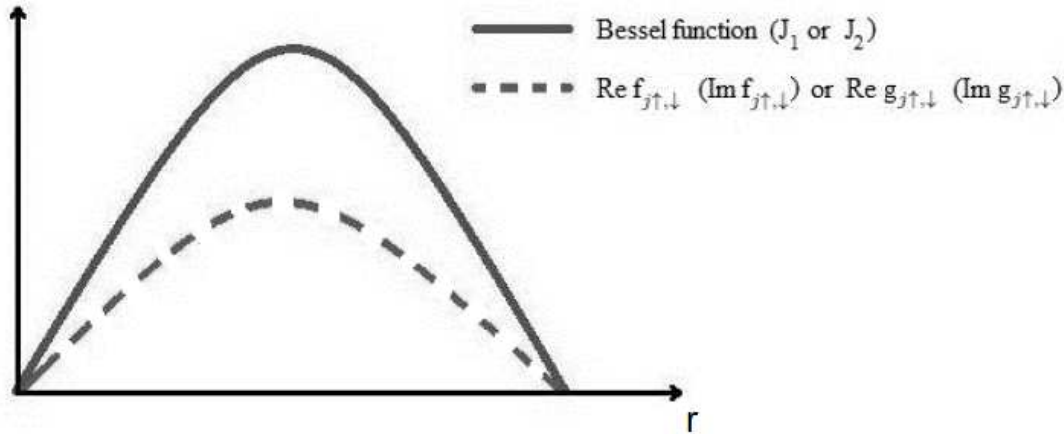}
\caption{Comparison of the real or imaginary part of
$f_{j\uparrow,\downarrow}$ ($g_{j\uparrow,\downarrow}$) and of the
corresponding Bessel function. Here, the $r$ coordinate denotes the distance from a tip. We have a boundary of a nanocone at $r_{0}=20$. We have zero probability of
finding an electron at a boundary.} \label{transform}
\end{figure}

Now we want to prove the correspondence between the solution of
(\ref{syst}) and the Bessel functions. But first, we need to change the scale
of the corresponding real or imaginary part of
$f_{j\uparrow,\downarrow}$ or $g_{j\uparrow,\downarrow}$ such
that the null points of the given function and the corresponding
Bessel function correlate (see Fig. \ref{transform}). Then, we will
do the described comparison for the case of $1$ defect and $j=2$ and
the corresponding Bessel functions $J_1$ and $J_2$ (see Table
\ref{table1}). The re--scaling we do numerically. For the value $E=0.75$ and the
unnormalized case, the re--scaled form of the solution of
(\ref{syst}) has the form

\begin{equation}f^r_{2\uparrow}(r)=e^{\frac{\alpha}{\frac{5.1103}{3.8317}r}+\beta \frac{5.1103}{3.8317}r}\sum\limits_{k=0}^{\infty}a_k\left(\frac{5.1103}{3.8317}r\right)^{\xi+k},\hspace{1cm}
f^r_{2\downarrow}(r)=e^{\frac{\alpha}{\frac{5.1071}{3.8317}r}+\beta \frac{5.1071}{3.8317}r}\sum\limits_{k=0}^{\infty}b_k\left(\frac{5.1071}{3.8317}r\right)^{\xi+k},\end{equation}
\begin{equation}g^r_{2\uparrow}(r)=e^{\frac{\alpha}{\frac{6.8490}{5.1356}r}+\beta\frac{6.8490}{5.1356}r}
\sum\limits_{k=0}^{\infty}c_k\left(\frac{6.8490}{5.1356}r\right)^{\xi_1+k},\hspace{1cm}
g^r_{2\downarrow}(r)=e^{\frac{\alpha}{\frac{6.8456}{5.1356}r}+\beta
\frac{6.8456}{5.1356}r}\sum\limits_{k=0}^{\infty}d_k\left(\frac{6.8456}{5.1356}r\right)^{\xi_1+k}.\end{equation}
We can do a shortcut of these expressions into the first 10 members
of the expansion. Here, we choose $\alpha=0$ and $\beta=1$.
Recalling that for the given values of the parameters,
$\xi_1=1.99987\doteq 2$, we have
\begin{equation}
{\rm Re}\,f^r_{2\uparrow}(r)=f^r_{2\uparrow}(r)\doteq 1.33 r - 0.17 r^3 + 6.95\cdot 10^{-3}r^5 - 1.45\cdot 10^{-4}r^7 + 1.81\cdot 10^{-6}r^9+\mathcal{O}(r^{10}),
\end{equation}
\begin{equation}
{\rm Im}\,f^r_{2\downarrow}(r)=f^r_{2\downarrow}(r)\doteq -1.33182 r + 0.17 r^3 - 6.93\cdot 10^{-3}r^5 + 1.44\cdot 10^{-4}r^7 - 1.80\cdot 10^{-6}r^9+\mathcal{O}(r^{10}),
\end{equation}
\begin{equation}
{\rm Re}g^r_{2\uparrow}(r)=g^r_{2\uparrow}(r)\doteq 0.33 r^2 - 2.78\cdot 10^{-2}r^4 + 8.69\cdot 10^{-4}r^6 - 1.45\cdot 10^{-5}r^8 + 1.51\cdot 10^{-7}r^{10}+\mathcal{O}(r^{11}),
\end{equation}
\begin{equation}
{\rm Im}g^r_{2\downarrow}(r)=g^r_{2\downarrow}(r)\doteq -0.33 r^2 + 2.77\cdot 10^{-2}r^4  - 8.67\cdot 10^{-4}r^6 + 1.45\cdot 10^{-5}r^8 - 1.50\cdot 10^{-7}r^10+\mathcal{O}(r^{11}).
\end{equation}
The corresponding expansions for the Bessel functions $J_1$, $J_2$ have the form
\begin{equation}J_1(r)\doteq 0.5r - 0.06r^3 + 2.60\cdot 10^{-3}r^5 - 5.43\cdot 10^{-5}r^7 + 6.78\cdot 10^{-7}r^9+\mathcal{O}(r^{10}),\end{equation}
\begin{equation}J_2(r)\doteq 0.125r^2 - 0.01r^4 + 3.26\cdot 10^{-4}r^6 - 5.43\cdot 10^{-6}r^8 + 5.65\cdot 10^{-8}r^10+\mathcal{O}(r^{11}).\end{equation}
Now, if we make a comparison of the coefficients corresponding to
the power series of Re$\,f^r_{2\uparrow}$, Im$\,f^r_{2\downarrow}$,
respectively, with the coefficients corresponding to $J_1$ and of the
coefficients corresponding to the power series of
Re$\,g^r_{2\uparrow}$, Im$\,g^r_{2\downarrow}$, respectively, with the
coefficients corresponding to $J_2$, we find that more or less, for
a concrete pair of functions, the ratio of the coefficients remains
constant and it approaches these values:
\begin{equation}\frac{{\rm Re}\,f^r_{2\uparrow}(r)}{J_1(r)}\doteq 2.67,\quad\frac{{\rm Im}\,f^r_{2\downarrow}(r)}{J_1(r)}\doteq -2.66,\quad\frac{{\rm Re}\,g^r_{2\uparrow}(r)}{J_2(r)}\doteq 2.67,\quad\frac{{\rm Im}\,g^r_{2\downarrow}(r)}{J_2(r)}\doteq -2.66.\end{equation}

In Table \ref{table1}, we see the concrete forms of the Bessel functions
which correspond to the case $N_{d}=1$ defect and different values
of the parameter $j$. However, the higher the value of $j$ is, the
more spread this correspondence is. On the whole, we can say that
for the case of $N_{d}=1$ defect and values of $j\geq 2$, the
analytical expression for the solution can be approximated as
\begin{equation}{\rm Re}\,f_{j\uparrow}\sim J_{j-1}(Er),\hspace{5mm}{\rm Im}\,f_{j\downarrow}\sim J_{j-1}(Er),\hspace{5mm}
{\rm Re}\,g_{j\uparrow}\sim J_{j}(Er),\hspace{5mm}{\rm
Im}\,g_{j\downarrow}\sim J_{j}(Er).\end{equation}

\begin{table}

\caption{Correspondence of the Bessel function with solutions of
(\ref{syst}) for $N_{d}=1$ defect and different values of $j$.}
\begin{tabular*}{0.55\textwidth}{@{\extracolsep{1cm}}ccccccc}
\hline & $j$ & Re$\,f_{j\uparrow}$ & Im$\,f_{j\downarrow}$ & Re$\,g_{j\uparrow}$ & Im$\,g_{j\downarrow}$ &
\\ \hline & $0$ & $-Y_2$ & $-Y_2$ & $Y_1$ & $Y_1$ &
\\ & $1$ & $-Y_1$ & $-Y_1$ & $J_1$ & $J_1$ &
\\ & $2$ & $J_1$ & $J_1$ & $J_2$ & $J_2$ &
\\ & $3$ & $J_2$ & $J_2$ & $J_3$ & $J_3$ &
\\ & $4$ & $J_3$ & $J_3$ & $J_4$ & $J_4$ &
\\ \hline
\end{tabular*}\label{table1}

\end{table}
 For the case of
$N_{d}=2$ and $N_{d}=3$ defects, the corresponding tables would be
more complicated, and the solution would be a linear combination of
Bessel functions of different kinds.

\end{document}